\documentclass[superscriptaddress,aps,prl,twocolumn,showpacs,preprintnumbers,amsmath,amssymb,floatfix,10pt]{revtex4-1}
\usepackage{graphicx}
\usepackage{dcolumn}
\usepackage{bm}
\usepackage{color}

\usepackage{braket}

\def\d{\downarrow}
\def\u{\uparrow}

\def\ba{\begin{eqnarray}}
\def\ea{\end{eqnarray}}
\def\beq{\begin{equation}}
\def\eeq{\end{equation}}

\begin{document}

\title{Kitaev honeycomb and other exotic spin models with polar molecules}


\author{Alexey V. Gorshkov}
\affiliation{Institute for Quantum Information \& Matter, California Institute of Technology, Pasadena, CA 91125, USA}
\author{Kaden R. A. Hazzard}
\affiliation{JILA, NIST and Department of Physics, University of Colorado, Boulder, CO 80309, USA}
\affiliation{Kavli Institute for Theoretical Physics, University of California, Santa Barbara, CA 93106, USA}
\author{Ana Maria Rey}
\affiliation{JILA, NIST and Department of Physics, University of Colorado, Boulder, CO 80309, USA}

\date{\today}

\begin{abstract}

We show that ultracold polar molecules pinned in an optical lattice can be used to access a variety of exotic spin models, including the Kitaev honeycomb model. Treating each molecule as a rigid rotor, we use DC electric and microwave fields to define superpositions of rotational levels as effective spin degrees of freedom, while dipole-dipole interactions give rise to interactions between the spins. In particular, we show that, with sufficient microwave control, the interaction between two spins can be written as a sum of five independently controllable Hamiltonian terms proportional to the five rank-2 spherical harmonics $Y_{2,q}(\theta,\phi)$, where $(\theta,\phi)$ are the spherical coordinates of the vector connecting the two molecules. 
To demonstrate the potential of this approach beyond the simplest examples studied in [S.\ R.\ Manmana \textit{et al.}, arXiv:1210.5518v2],
we focus on the realization of the Kitaev honeycomb model, which can support exotic non-Abelian anyonic excitations. We also discuss the possibility of generating spin Hamiltonians with arbitrary spin $S$, including those exhibiting SU($N$=$2S$+$1$) symmetry.

\end{abstract}


\maketitle




\section{Introduction}

Recent experimental progress in the control of ultracold polar molecules  \cite{ni08, aikawa10,deiglmayr08,ospelkaus10,demiranda11,chotia12} has  stimulated the study of many-body systems featuring strong dipole-dipole interactions \cite{baranov08,lahaye09,trefzger11,baranov12}. A particularly promising research area \cite{brennen07,buchler07b,wall09,wall10,schachenmayer10,perezrios10,herrera10,kestner11,zhou11,dalmonte11b,hazzard12b,lemeshko12,sowinski12,maik12} of implementing lattice Hamiltonians from dipole-dipole-interacting molecules has emerged following the pioneering works of Refs.\ \cite{barnett06,micheli06}. 
This research area largely owes its promise to the great degree of controllability that external DC, microwave, and optical fields can provide over rotational levels and dipole-dipole interactions \cite{micheli06, brennen07,buchler07,buchler07b,micheli07,gorshkov08c,lin10b,cooper09,wall09,wall10,schachenmayer10,kestner11,lemeshko11}. In Ref.\ \cite{manmana12}, we demonstrated that this controllability can yield a variety of spin Hamiltonians, in which spin states are encoded in superpositions of rotational states of the molecules. While dipole-dipole interactions are often thought of as being proportional to $1-3 \cos^2 \theta$, where $\mathbf{R} = (R,\theta,\phi)$ are the spherical coordinates of the vector connecting the two dipoles, we showed that not only the overall amplitude and sign of the spin Hamiltonian but also its individual terms 
can depend on $\hat{\mathbf{R}}$. 
We harnessed this dependence to propose the realization of spin models known to harbor topological phases of matter \cite{manmana12}. 

In contrast to Ref.\ \cite{manmana12}, which focused on the simplest examples, in the present paper we discuss a range of models that are more interesting but harder to implement experimentally. 
Specifically, as in Ref.\ \cite{manmana12}, we emphasize that the interaction Hamiltonian between two molecules can, in principle, be a sum of five independently controllable terms proportional to the five rank-2 spherical harmonics $Y_{2,q}(\theta,\phi)$. 

We then study a specific class of Hamiltonians with an arbitrary spin $S$. This class of Hamiltonians can be used to realize \cite{manmana12} the general SU(2) symmetric $S = 1$ model, i.e.\ the  bilinear-biquadratic spin model, which has a rich phase diagram even in one dimension  \cite{schollwock96,garcia-ripoll04,brennen07}.  It can also give rise to SU($N$=$2S$+$1$) spin models, which have recently attracted a great deal of theoretical and experimental attention due to their potential to access exotic strongly-correlated phases in alkaline-earth atoms \cite{gorshkov10,cazalilla09,taie10,daley11}. The strong dipole-dipole interactions of the polar-molecule implementation of SU($N$) models can be larger than the SU($N$)-symmetric superexchange interactions in cold-atom implementations \cite{gorshkov10}.

Then we turn to the primary example of this paper: the implementation of the Kitaev honeycomb model. In contrast to the proposal of Ref.\ \cite{micheli07}, our implementation of the Kitaev honeycomb model relies on direct -- rather than perturbative -- dipole-dipole interaction and hence gives rise to stronger interactions, which are easier to access experimentally. Other proposals for implementing the Kitaev honeycomb model use solid-state Mott insulators \cite{jackeli09}, trapped ions \cite{schmied11}, superconducting quantum circuits \cite{you10}, weak superexchange interaction between cold atoms in optical lattices \cite{duan03}, and an experimentally challenging system of coupled cavity arrays \cite{xiang12}. We end the manuscript with a discussion of how we use microwave fields to create the dressed states that enable our implementations, followed by an outlook. 


\section{Setup}

\begin{figure}[t]
\begin{center}
\includegraphics[width = 0.99 \columnwidth]{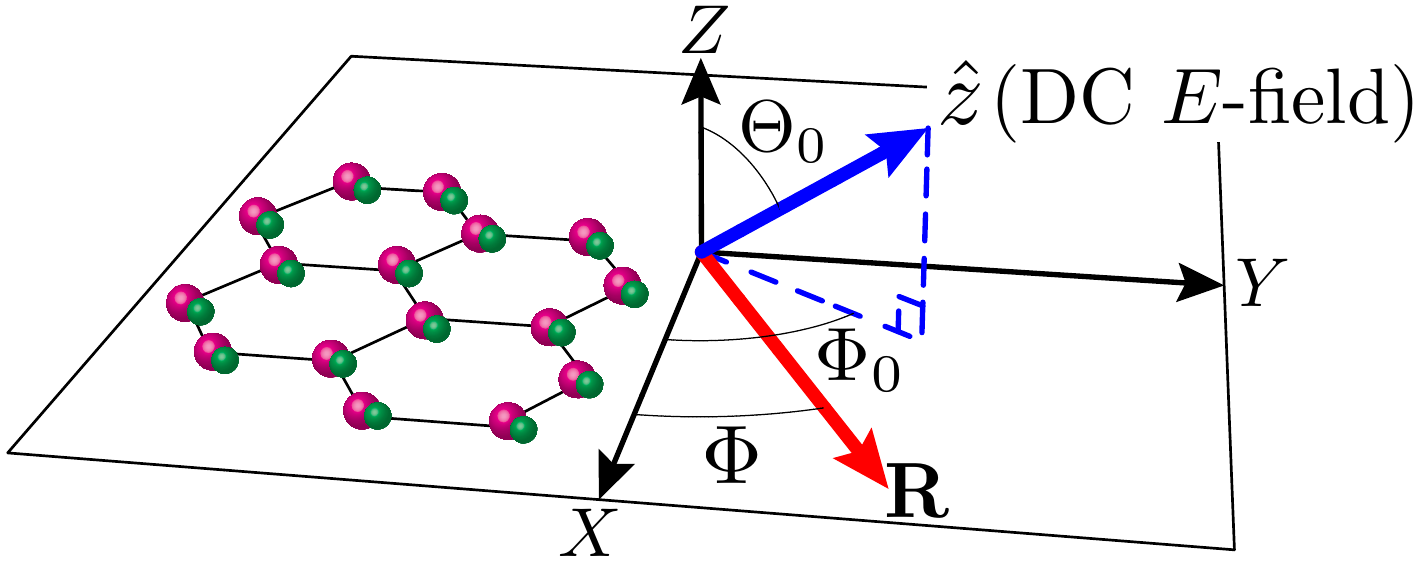}
\caption{(color online). 
The lattice of molecules is in the $XY$ plane. The direction of the DC electric field is $\hat z$. The $xyz$ coordinate system is obtained from the $XYZ$ coordinate system by rotating the former around $\hat Z$ by $\Phi_0$ and then rotating it around $\hat y$ by $\Theta_0$. A typical vector $\mathbf{R}$ with spherical coordinates  $(R,\theta, \phi)$ in the $xyz$ coordinate system has polar coordinates $(R,\Phi)$  in the $XY$ plane. \label{fig:scheme}}
\end{center}
\end{figure}

Consider a deep optical lattice in the $XY$ plane shown in Fig.\ \ref{fig:scheme} and loaded with one polar molecule per site. Each molecule is described by a rigid rotor with rotational angular momentum operator $\mathbf{N}$, rotational constant $B$, and dipole moment operator $\mathbf{d}$. In the presence of an electric field $E$  along $\hat z$, each molecule is described by the Hamiltonian
\ba
H_0 = B \mathbf{N}^2 - E d^z.
\ea
In general, the molecules can be allowed to hop to give rise to Hubbard-type \cite{barnett06,he11} or $t$-$J$-type \cite{gorshkov11b,gorshkov11c} models with highly tunable anisotropic long-range spin-spin interactions. In this paper, however, we will assume that the lattice is so deep that tunneling is negligible and molecules are pinned in the motional ground state on each site. The field-free (i.e.~$E=0$) eigenstates of $H_0$ are the simultaneous eigenstates of $\mathbf{N}^2$ and $N_z$ with eigenvalues $N (N+1)$ and $M$, respectively. Let us denote with $\ket{N,M}$ the eigenstates of $H_0$, which the field-free eigenstates connect to as $E$ is turned on.  
For each molecule, we define \cite{micheli07} $d^\pm =\mp (d^x \pm i d^y)$, which changes the $M$ of the molecule by $\pm 1$, while $d^0 = d^z$ couples rotational states with the same $M$. There are no selection rules on $N$ for $E \neq 0$.

Consider two molecules $i$ and $j$ separated by $\mathbf{R}_{ij}$, which has polar coordinates $(R_{ij},\Theta_{ij})$ in the $XY$ plane and spherical coordinates $(R_{ij},\theta_{ij},\phi_{ij})$ in the $xyz$ coordinate system. The dipole-dipole interaction between these two molecules is \cite{brown03}
\ba
H_{ij} &=& \frac{\mathbf{d_i} \cdot \mathbf{d_j} - 3 (\mathbf{d_i} \cdot \hat{\mathbf{R}}_{ij}) (\mathbf{d_j} \cdot \hat{\mathbf{R}}_{ij})}{R_{ij}^3} \\
&=&- \frac{\sqrt{6}}{R_{ij}^3} \sum_{q = -2}^{2} (-1)^q C_{-q}^2(\theta_{ij}, \phi_{ij}) T^2_q(\mathbf{d_i}, \mathbf{d_j}), \label{eq:dd}
\ea
where $C^k_q(\theta, \phi) = \sqrt{\frac{4 \pi}{2 k + 1}} Y_{kq}(\theta,\phi)$, $Y_{kp}$ are spherical harmonics,
$T^2_{\pm 2}  = d_i^\pm  d_j^\pm$, $T^2_{\pm 1}  = \left(d_i^0  d_j^\pm + d_i^\pm  d_j^0\right)/\sqrt{2}$, and 
$T^2_0= \left(d_i^-  d_j^+ + 2 d_i^0  d_j^0 + d_i^+  d_j^-\right)/\sqrt{6}$, so that $T^2_q$ changes the total $M$ of the two molecules by $q$. $T^2_{\pm 1}$ contributes only at those values of the DC electric field that have $\sigma^\pm$ ($M \rightarrow M\pm 1$) and $\pi$ ($M$-conserving) transitions of matching frequency.

In each molecule, we couple states $\ket{N,M}$ with several microwave fields and choose $2 S + 1$ dressed states (linear combinations of states $\ket{N,M}$) to define an effective spin $S$ system. We will focus in this paper on homogeneous driving, which is easier to achieve as it can be done with microwave -- as opposed to optical -- fields. Extensions to inhomogeneous driving allow for an even richer class of Hamiltonians \cite{yao12d, yao12e}.  Assuming dipole-dipole interactions are too weak to take the molecules out of the $2 S + 1$ chosen dressed states, we can project dipole-dipole interactions onto these states to obtain a spin-$S$ interaction Hamiltonian of the form $H = \frac{1}{2} \sum_{i \neq j} H_{ij}$, where
\ba
R_{ij}^3 H_{ij} = \mathbf{v}(\theta_{ij},\phi_{ij}) \cdot \mathbf{H}. \label{eq:gen}
\ea
Here 
\ba
\mathbf{v}(\theta,\phi) = \left(-2 C^2_0, - \sqrt{6} \textrm{Re}[C^2_2], \sqrt{6} \textrm{Im}[C^2_2], \textrm{Re}[C^2_1], \textrm{Im}[C^2_1]\right) \nonumber
\ea
is a real five-component vector describing the five different angular dependences (the prefactors are chosen for later convenience). 
Each of the five components of $\mathbf{H}$ is a Hamiltonian acting on the Hilbert space of the two spin-$S$ particles $i$ and $j$ and comes with its own angular dependence. Specific examples for the components of $\mathbf{H}$ are given below, for example Eq.\ (\ref{Hgen}) for the spin-$1/2$ case.
Due to Hermiticity and symmetry under the exchange of the two particles, each component of $\mathbf{H}$ has $[(2 S + 1)^4 + (2 S + 1)^2]/2$ independent real coefficients.  With an appropriate choice of rotational states, DC electric field strength, and a sufficient number of microwave fields, one might envision achieving full control over all $5 [(2 S + 1)^4 + (2 S + 1)^2]/2$ coefficients, which, together with $\Theta_0$, $\Phi_0$, and a choice of lattice, determine the system. Requiring in addition that the total number of particles in any given internal state is conserved gives only $(2 S + 1)^2$ independent coefficients in each angular dependence.

Eq.\ (\ref{eq:gen}) allows one to access a great variety of exotic spin Hamiltonians. 
Since we will discuss $S = 1/2$ examples below in Eq.\ (\ref{Hgen}), here we only point out that, 
in Ref.\ \cite{manmana12}, we showed how Eq.\ (\ref{eq:gen}) can be used to realize the most general SU(2)-symmetric $S = 1$ Hamiltonian (the so-called bilinear-biquadratic Hamiltonian) restricted to the $C^2_0$ angular dependence. in Ref.\ \cite{manmana12}, we 
also briefly mentioned the possibility of realizing the Kitaev honeycomb model. In the present paper, we 
discuss the details behind the realization of the Kitaev honeycomb model and 
provide additional insights into how to generate spin Hamiltonians with an arbitrary $S$.

\section{Spin Hamiltonians with arbitrary $S$}

In this section, we show how to obtain a variety of spin Hamiltonians with an arbitrary $S$. For simplicity, in each molecule, we choose $Q$ distinct $|N,M\rangle$ states and label them as $|a\rangle$, where $a = 1, \dots, Q$. We break this set of $Q$ states into $2 S + 1$ disjoint sets labeled by $p = - S, \dots, S$. We couple the states within each set with microwave fields to form dressed states in the rotating frame. We will show in detail below that $\sim n$ microwave fields are needed to couple $n$ states and to create any desired linear combination out of them. We then choose one dressed state from each set to create the single-spin basis $|p\rangle = \sum_{a(p)} \sqrt{x_a} |a\rangle$ with rotating frame energies $E_p$. Here $\sum_{a(p)}$ 
means that $a$ is summed over the states belonging to the set $p$. The coefficients $x_a$ are assumed to be nonnegative real numbers: allowing for superpositions with complex coefficients does not allow any additional tunability in this example \footnote{Allowing for complex coefficients does not allow any additional tunability in this example because we assume that,  in terms of the bare non-microwave-dressed states, the two-molecule state $\ket{ab}$ is connected via $H_{ij}$ only to itself and to $\ket{ba}$.}. 
The states $|p\rangle$ and $|q\rangle$ will refer to dressed states, while states $\ket{a}$ and $\ket{b}$ will refer to bare states.
(One could also choose fewer sets and choose more than one dressed state from the same set.)
For simplicity, we further assume that dipole-dipole interactions are so weak and the states are chosen in such a way that the two-molecule state $|p q\rangle$ is connected via $H_{ij}$ only to itself and to $|q p\rangle$, while all the other processes are off-resonant and are negligible. Similarly, now in terms of the bare non-microwave-dressed states, we assume that the states are chosen in such a way that the two-molecule state $|ab\rangle$ is connected via $H_{ij}$ only to itself and to $|ba\rangle$, while all the other processes are off-resonant and are negligible. Finally, also for simplicity, we assume  that the states are chosen in such a way that only $C^2_0$ contributes (an extension that includes the other four components of $\mathbf{v}$ is straightforward). The assumptions in the previous three sentences are generically satisfied if, for example, all the states $|a\rangle$ involved have $M \geq 0$ and no accidental degeneracies occur. 
We then have
\ba
H_{ij}
&=&  \frac{1- 3 \cos^2 \theta_{ij}}{R_{ij}^3}  \Bigg[\sum_p B_p |p p\rangle \langle p p| + \sum_{p,q} A_p A_q |p q\rangle \langle p q| \nonumber \\
&& + \sum_{p < q} \frac{J_{p,q}}{2} (|q p\rangle \langle p q| + \textrm{h.c.})\Bigg], \label{Hdds1}
\ea
where $A_p = \sum_{a(p)} x_{a} \mu_{a}$, $B_p = \sum_{a(p) < b(p)} x_{a} x_{b} d_{a,b}$, $J_{p,q} = \sum_{a(p),b(q)} x_{a} x_{b} d_{a,b}$, where
\ba
d_{a, b} = \left\{\begin{array}{cl} 2 \mu_{a,b}^2 &\textrm{ if } M(a) = M(b) \\
- \mu_{a,b}^2 & \textrm{ if } |M(a)-M(b)| = 1 \\
0 & \textrm{ otherwise} \end{array} \right.. \label{eq:dij}
\ea
Here $\mu_{a,b} = \langle a|d^{M(a)-M(b)} |b\rangle$ and $\mu_{a} = \langle a|d^0|a\rangle$. 
$A_p$ can be regarded as an effective dipole moment of state $\ket{p}$, while $B_p$ encompasses the contribution to the interaction from the transition dipole moments between the states $a(p)$. On the other hand, $J_{p,q}$ describes the flip-flop transition involving states $\ket{p}$ and $\ket{q}$ and is driven by transition dipole moments coupling $a(p)$ to $b(q)$. The factor of $2$ and the minus sign in Eq.\ (\ref{eq:dij}) arise from the fact that two identical dipoles rotating in phase in the $x$-$y$ plane experience, on average, an interaction equal to negative one-half of that felt by these dipoles had they been pointing along $\hat{\mathbf{z}}$ \cite{gorshkov08c}.

The implementation of  SU($N$=$2S$+$1$)-symmetric models   
is an  attractive case to consider given that  such models have great potential for generating exotic phases including chiral spin liquids \cite{hermele09,hermele11}.
The expression in the square brackets in Eq.\ (\ref{Hdds1}) possess SU($N$=$2S$+$1$) symmetry if and only if it is proportional to $\sum_{p,q} |pq\rangle \langle qp|$ up to an additive constant. Necessary and sufficient conditions for this
 are that $A_p A_q$, $J_{p,q}$, and $B_p + A_p^2$ do not depend on $p$ and $q$ for all $p < q$ and that $B_p + A_p^2 = A_p A_q + J_{p,q}/2$ for all $p < q$. As an example, for $N = 3$, it is sufficient to satisfy $A_1 = A_2 = 0$ and $B_1   =  B_2 = B_3 + A_3^2 = \tfrac{1}{2} J_{1,2} = \tfrac{1}{2} J_{1,3} = \tfrac{1}{2}  J_{2,3}$. In Ref.\ \cite{manmana12}, we showed how to realize this and, in fact, an arbitrary SU(2) symmetric $S = 1$ interaction. Finding specific level configurations for higher $N$ is postponed until future work.



\section{The general $S = 1/2$ Hamiltonian} 

The most general $S=1/2$ Hamiltonian implementable with dipole-dipole interactions is 
\ba
R_{ij}^3 H_{ij} &=& \mathbf{v}(\theta_{ij},\phi_{ij}) \cdot [\mathbf{V} n_i n_j  + \mathbf{W}_x (n_i S^x_j + S^x_i n_j)   \nonumber \\
&&+ \mathbf{W}_y (n_i S^y_j + S^y_i n_j) + \mathbf{W}_z (n_i S^z_j + S^z_i n_j)  \nonumber \\
&&+ \mathbf{J}_{xx} S^x_i S^x_j  + \mathbf{J}_{yy} S^y_i S^y_j+ \mathbf{J}_{zz} S^z_i S^z_j   \nonumber \\
&& + \mathbf{J}_{xy} (S^x_i S^y_j + S^y_i S^x_j) + \mathbf{J}_{xz} (S^x_i S^z_j + S^z_i S^x_j) \nonumber \\
&&+ \mathbf{J}_{yz} (S^y_i S^z_j + S^z_i S^y_j)],\label{Hgen}
\ea
where we have 10 real 5-component vectors in the square brackets.  By analogy with Refs.\ \cite{gorshkov11b,gorshkov11c,manmana12}, we expect that these vectors might be independently controllable, an issue whose details we leave for future investigation.
 For this, we need $\sim 50$ microwave fields. In Eq.\ (\ref{Hgen}), $n_i$ is the occupation of site $i$. In the present manuscript, $n_i = 1$ for all sites. However, this more general form of the Hamiltonian is necessary when hopping is allowed, such as in Refs.\ \cite{barnett06,he11,gorshkov11b,gorshkov11c}.  

In general, the interaction in Eq.\ (\ref{Hgen}) is also accompanied by on-site Hamiltonian terms. In particular, if the on-site energy difference between $\ket{\u}$ and $\ket{\d}$ (the two dressed basis states making up the spin-1/2) is much larger than the strength of $H_{ij}$, only those interaction terms that conserve the total $S^z$ play a role \cite{gorshkov11b,gorshkov11c}. In order to realize models such as the quantum compass model \cite{kugel73}, the Kitaev honeycomb model \cite{kitaev06}, and the Kitaev quantum double models \cite{kitaev03} implemented with two-body interactions \cite{brell11}, we need to break $S^z$ conservation and realize terms, such as $S^x_i S^x_j$. 
To do this, we simply tune $\ket{\u}$ and $\ket{\d}$ to be degenerate. 

For $n_i = 1$, the Hamiltonian in Eq.\ (\ref{Hgen}) is the special case of Eq.\ (\ref{eq:gen}): its restriction to $S = 1/2$. Despite the restriction to $S = 1/2$, the Hamiltonian in Eq.\ (\ref{Hgen}) is still extremely powerful, as illustrated by the following examples. First, the commonly used density-density interaction with the $1- 3 \cos^2 \theta_{ij}$ angular dependence (see e.g.~Refs.\ \cite{baranov08,trefzger11}) comes from the first component of $\mathbf{V}$, while density-density interactions with more exotic angular dependences can be obtained by making use of all five components of $\mathbf{V}$. Second, the first components of  $\mathbf{J}_{xy}$, $\mathbf{J}_{zz}$, $\mathbf{W}_z$, and $\mathbf{V}$  give rise to the spin-spin, density-spin, and density-density interactions of the generalized $t$-$J$ model (referred to as the $t$-$J$-$V$-$W$ model) of Refs.\ \cite{gorshkov11b,gorshkov11c}. Third, the first three components of $\mathbf{J}_{xy}$ and $\mathbf{J}_{zz}$ give rise to an XXZ model featuring a direction-dependent spin-anisotropy \cite{manmana12}. In addition to providing access to symmetry protected topological phases in spin ladders \cite{manmana12}, such an XXZ model can also be used to obtain a variety of exotic antiferromagnets including a square-lattice Heisenberg model, in which 
the ratio between coupling strengths on $\hat X$ and $\hat Y$ bonds is tunable \cite{kim:monte_2000}. Fourth, the $\mathbf{J}_{xx}$ and $\mathbf{J}_{zz}$ terms can be used to realize the quantum compass model \cite{kugel73}. In the present paper, we demonstrate the power of Eq.\ (\ref{Hgen}) on another important example: the Kitaev honeycomb model \cite{kitaev06}.


\section{Kitaev honeycomb}

\begin{figure}[t]
\begin{center}
\includegraphics[width = 0.99 \columnwidth]{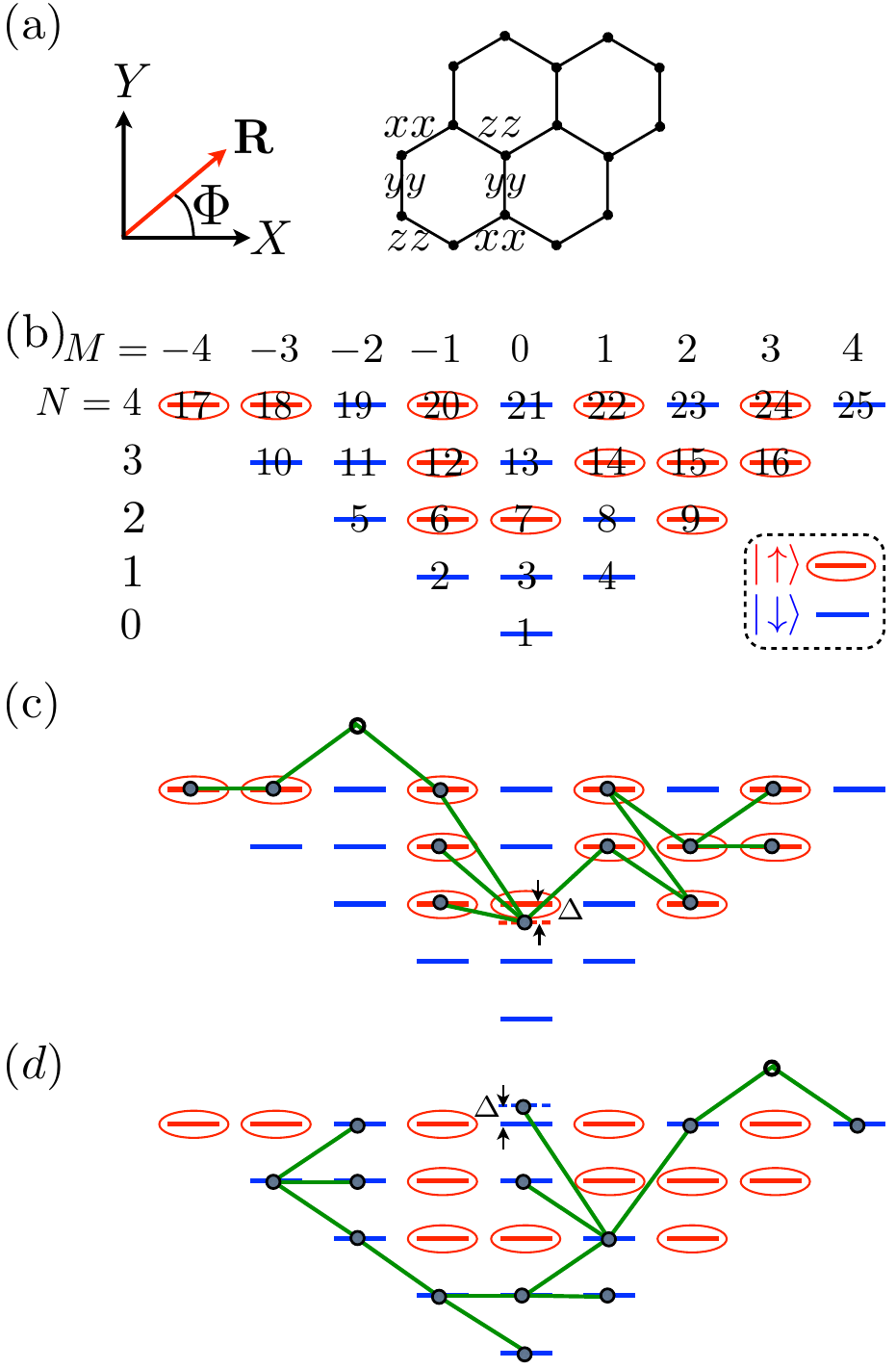}
\caption{Implementation of the Kitaev honeycomb model. (a) The honeycomb model: $S^x_i S^x_j$, $S^y_i S^y_j$, and $S^z_i S^z_j$ interactions along $\Phi_{ij} = \pi/6$, $\pi/2$, and $5 \pi/6$, respectively.  (b) The rotational levels used. The diagram is schematic: the real system is anharmonic and levels $|N,M\rangle$ with the same $N$ are non-degenerate (unless the levels have the same $|M|$). The dressed states are $\ket{\u}$ (linear combination of states indicated by ovals) and $\ket{\d}$ (the rest). (c) Microwave fields that can be used to create the dressed state $\ket{\u}$. Notice that level  $|2,0\rangle$ is used off-resonantly with a detuning $\Delta$. (d)   Microwave fields that can be used to create the dressed state $\ket{\d}$. Notice that level  $|4,0\rangle$ is used off-resonantly with a detuning $\Delta$. \label{fig:mw}}
\end{center}
\end{figure}

In Fig.\ \ref{fig:scheme} and in the inset to Fig.\ \ref{fig:mw}(a), we show the Kitaev honeycomb lattice with  Hamiltonian \cite{kitaev06}
\ba
\!\!\!\! H = J_{xx} \!\!\! \sum_{x-\textrm{link}}  S^x_i  S^x_j +  J_{yy} \!\!\! \sum_{y-\textrm{link}}  S^y_i S^y_j +  J_{zz} \!\!\! \sum_{z-\textrm{link}}  S^z_i S^z_j. 
\ea
We are interested in phase B, which contains the point $J_{xx} = J_{yy} = J_{zz} < 0$. This phase is gapless, but, in the presence of a magnetic field, 
\ba
H_B = \sum_j (B_x S^x_j + B_y S^y_j + B_z S^z_j), \label{eq:B}
\ea
with all three components $B_x$, $B_y$, and $B_z$ being nonzero, acquires a gap and supports non-Abelian anyonic excitations. Non-Abelian anyons have quantum computing applications and can be used, for example,  for topologically protected quantum state transfer \cite{yao11c}.


We use Eq.\ (\ref{Hgen}) for a system of $n=1$ molecules per site on a honeycomb optical lattice --  which can be implemented with three 
laser beams \cite{grynberg93,wunsch08,tarruell12}  -- to implement the Kitaev honeycomb model. Specifically, we would like to obtain $S_i^x S_j^x$ interactions  along $\Phi_{ij} = \pi/6$, $S^y_i S^y_j$ interactions along $\Phi_{ij} =  \pi/2$, and $S^z_i S^z_j$ interaction along $\Phi_{ij} = 5 \pi/6$ [see Fig.\ \ref{fig:mw}(a)]. It is worth noting that interactions along $\Phi_{ij}$ and $\Phi_{ij}+\pi$ are  the same. 

Let us first point out some general facts that hold independently of the direction and magnitude of the applied DC electric field and of the dressed states $\ket{\u}$ and $\ket{\d}$ that we choose. The terms in Eq.\ (\ref{Hgen}) involving $\mathbf{V}$ play no role since they are proportional to identity. 
Ignoring boundary effects, the terms in Eq.\ (\ref{Hgen}) involving $\mathbf{W}_x$, $\mathbf{W}_y$, and $\mathbf{W}_z$ give rise to a uniform magnetic field, precisely as required by Eq.\ (\ref{eq:B}); we will compute this field below. 

Having taken care of the $\mathbf{V}$ and $\mathbf{W}_\alpha$ terms, we omit them from the Hamiltonian. We would like the remaining terms in the Hamiltonian ($\mathbf{J}_{xx}, \mathbf{J}_{yy}, \mathbf{J}_{zz}, \mathbf{J}_{xy}, \mathbf{J}_{xz}, \mathbf{J}_{yz}$) to satisfy
\ba
H_{ij}(\left\{\tfrac{\pi}{6},\tfrac{\pi}{2},\tfrac{5 \pi}{6}\right\}) = - \frac{J}{R_{ij}^3} \left\{S^x_i S^x_j,S^y_i S^y_j,S^z_i S^z_j\right\} \label{eq:3hs} 
\ea
 with $J > 0$ for the three indicated values of $\Phi_{ij}$.  Since, for any $\Phi_{ij}$, $\mathbf{v}(\Phi_{ij})$ can be written as a linear combination of $\mathbf{v}(\pi/6)$, $\mathbf{v}(\pi/2)$, and $\mathbf{v}(5 \pi/6)$, Eq.\ (\ref{eq:3hs}) completely determines the Hamiltonian at all $\Phi_{ij}$:
 \ba
 H_{ij}(\Phi_{ij}) &=& \sum_{n=1}^3  \frac{1- 2 \cos\left(2 \Phi_{ij} - 2 (n+1) \pi/3\right)}{3} \nonumber \\
 &&\times H_{ij}\left( \tfrac{(2 n-1) \pi}{6}\right) \label{eq:all} \\
 &=& - \frac{J}{3 R_{ij}^3} \Bigg\{\left[1- 2 \cos\left( 2 \Phi_{ij} - 4 \pi/3\right)\right] S^x_i S^x_j  \nonumber  \\
&& \quad \quad \quad  + \left[1- 2 \cos\left( 2 \Phi_{ij}\right)\right] S^y_i S^y_j   \nonumber \\
&&  \quad \quad \quad  + \left[1- 2 \cos\left( 2 \Phi_{ij} - 2 \pi/3\right)\right] S^z_i S^z_j \Bigg\}. \nonumber
 \ea
We will show how to realize this momentarily. First, observe that, for example, along $\Phi_{ij}=\pi/3$, which is halfway between $S^x_i S^x_j$ and $S^y_i S^y_j$,
\ba
H_{ij}(\pi/3) = - \frac{J}{3 R_{ij}^3} (2 S^x_i S^x_j + 2 S^y_i S^y_j - S^z_i S^z_j).
\ea
Because of the $1/R^3$ dependence, the strength of such next-nearest-neighbor interactions (along $\Phi_{ij} = 0, \pi/3, 2 \pi/3$) is reduced relative to the nearest-neighbor interactions by $1/3^{3/2} \approx 1/5$, which will be the largest correction introduced by long-range interactions. While in some cases long-range corrections are weak enough to ensure the survival of the desired phases \cite{manmana12}, it is an open question whether this holds for the present example.

It now remains to find the DC electric field strength and direction, as well as the dressed states, that yield Eq.\ (\ref{eq:3hs}). We assume that the DC electric field is along the $Z$ axis (i.e.~$\Theta_0 = 0$) and  take $\Phi_0 = 0$. In that case, the last two components of $\mathbf{v}$ vanish, which means that $C^2_{\pm 1}$ terms do not contribute ($v_4 = v_5 = 0$). This is fine because those terms are the hardest to generate experimentally as they are nonvanishing only at specific values of the DC electric field, and it will turn out they are not required to generate the Kitaev Hamiltonian. At $\Theta_0 = \Phi_0 = 0$, the uniform magnetic field coming from $\mathbf{W}_\alpha$ terms is easy to calculate. Indeed, $v_2$ and $v_3$ do not contribute to the magnetic field since the honeycomb lattice is invariant under $2 \pi/3$ rotations and since $\sum_{n = 0}^2\{v_2(\Phi_1+n 2 \pi/3),v_3(\Phi_1+n 2 \pi/3)\}= \{0,0\}$ for any $\Phi_1$. Finally, summing over all bonds connected to a given site, the term $\mathbf{W} \cdot (\mathbf{S}_i+\mathbf{S}_j)$ with angular dependence $v_1$ gives rise to a uniform magnetic field of strength $6.58 \mathbf{W}$.

Since $v_4 = v_5 = 0$, we drop the last two components of $\mathbf{v}$ and $\mathbf{H}$ to obtain
\ba
\mathbf{v}(\Phi) =  (1, - \tfrac{3}{2} \cos(2 \Phi) ,  \tfrac{3}{2} \sin(2 \Phi))
\ea
and
\ba
\mathbf{H} &=& \{d_i^0  d_j^0 + \tfrac{1}{2} (d_i^-  d_j^+  + d_i^+  d_j^-), \nonumber \\
&& d_i^+  d_j^+ + d_i^-  d_j^-, i (d_i^+  d_j^+ - d_i^-  d_j^-)\}. \label{eq:Hbf}
\ea
The linear independence of $\mathbf{v}(\pi/6)$, $\mathbf{v}(\pi/2)$, and $\mathbf{v}(5 \pi/6)$  allows, in principle, for the possibility of obtaining Eq.\ (\ref{eq:3hs}).



We will use the 25 states shown and numbered in Fig.\ \ref{fig:mw}(b). 
We write the two dressed states as $|\sigma\rangle = \sum_{a(\sigma)} y_{a} |a\rangle$, where $y_a$ are complex, $\sum_{a(\sigma)} |y_a|^2 = 1$, $\sigma = \u, \d$, and $\sum_{a(\sigma)}$ means that one sums $a$ over the 12 (13) states belonging to $\ket{\u}$ ($\ket{\d}$) in Fig.~\ref{fig:mw}(b). 
We now keep only resonant terms 
and project $\mathbf{H}$ in Eq.\ (\ref{eq:Hbf}) onto dressed states $\ket{\u}$ and $\ket{\d}$. 

We work at a DC electric field $E = 10 B/d$. The following transitions contribute to $H_1$: $\ket{N, M} \ket{N',M'} \rightarrow \ket{N, M} \ket{N',M'}$, $\ket{N, M} \ket{N',M'} \rightarrow \ket{N', M'} \ket{N,M}$ (for $|M\!-\!M'| \leq 1$, as dictated by electric-dipole selection rules), and $\ket{N, M} \ket{N',-M} \rightarrow \ket{N', M} \ket{N,-M}$. The transition $\ket{N,-M\!\!-\!1} \ket{N',M} \rightarrow \ket{N',-M} \ket{N,M+1}$ is the only one contributing to $d^+_i d^+_j$ in $H_2$ and $H_3$, while $d^-_i d^-_j$ is the Hermitian conjugate of $d^+_i d^+_j$. Projecting these terms on $\ket{\u}$ and $\ket{\d}$, we obtain an expression for $\mathbf{H}$ in terms of $y_a$. This allows us to get an expression for $\mathbf{W}$ and for the left-hand side of Eq.\ (\ref{eq:3hs}). Since a magnetic field that is too large will eventually take us out of the Kitaev B phase, we first verify that we can obtain Eq.\ (\ref{eq:3hs}) with a minimal magnetic field $\mathbf{W}$. One can then easily verify that the set of $y_a$ given in footnote 
 \footnote{$\{y_1, \dots, y_{25}\} = \{-0.0016-0.0230 i,-0.1077+0.1635 i,0.2272+0.0323 i,0.0168-0.0024 i,-0.0277-0.1580 i,0.2258+0.1997 i,0.0503+0.0694 i,-0.4994+0.1819 i,0.3618-0.0554 i,0.3558+0.1635 i,0.0530-0.1025 i,0.1942-0.0515 i,-0.1135+0.0599 i,0.2284+0.2537 i,-0.3417-0.1720 i,-0.3946+0.0264 i,-0.0739+0.0092 i,-0.3268-0.0782 i,-0.0200+0.0088 i,-0.1105-0.0720 i,0.5198+0.0203 i,-0.0598-0.0452 i,-0.1838+0.1808 i,-0.3090-0.2669 i,-0.1163+0.2564 i\}.$} gives $J = 0.0284 d^2$ and $6.58 |\mathbf{W}| = 7 \times 10^{-7} d^2$ (which is negligibly small). 

By adjusting the coefficients $y_a$, one can find points with larger values of $|\mathbf{W}|$.  For example, the coefficients $y_a$ given in footnote \footnote{$\{y_1, \dots, y_{25}\} = \{0.0041-0.0220 i,-0.1773+0.1736 i,0.2159+0.0282 i,0.0212-0.0025 i,-0.0394-0.1639 i,0.2020+0.1610 i,0.0082+0.1014 i,-0.5091+0.1405 i,0.2452-0.0578 i,0.4216+0.1754 i,0.0255-0.0924 i,0.1975-0.0497 i,-0.1510+0.0547 i,0.2955+0.2858 i,-0.4215-0.1541 i,-0.4519+0.0592 i,-0.0477+0.0083 i,-0.3147-0.0823 i,-0.0449+0.0088 i,-0.0298-0.1000 i,0.4665+0.0034 i,-0.0981-0.0361 i,0.0425+0.1662 i,-0.1913-0.2707 i,-0.1028+0.2820 i\}$.} achieve 
$J = 0.0264 d^2$ and 
$6.58 \mathbf{W} = \{0.00072, 0.00084, -0.03937\}$. 

\section{Microwave dressing}


So far, we have not discussed in detail what exact microwave couplings are required to create the dressed states $\ket{\u}$ and $\ket{\d}$ needed to implement the Kitaev honeycomb model. In this Section, we provide such a discussion. The methods we present here are generally applicable to the creation of generic spin models from dipolar interactions. 

 Although generating large numbers of precisely tuned microwaves is experimentally feasible \cite{dian08}, the effort in obtaining and verifying the spectrum of microwaves increases with the number of frequencies.  It is thus experimentally desirable to use as few microwave fields as possible.  In order to create a dressed state out of $n$ bare states, one needs at least $n-1$ microwaves (assuming the same microwave cannot be used to couple more than one transition). Therefore, let us first address the question of whether $n-1$ microwaves suffice to prepare a dressed state at any prescribed  rotating frame energy 
 featuring any prescribed superposition of the bare states.

Suppose we would like to prepare a dressed state $|D\rangle = \sum_{a=1}^n y_a \ket{a}$ out of the $n$ bare states $\ket{a}$ ($a = 1, \dots, n$) with prescribed coefficients $y_a$ and a prescribed rotating-frame energy $E$ relative to the bare state $\ket{m}$. Suppose state $\ket{a}$ is coupled to state $\ket{a+1}$ using a microwave with Rabi frequency $\Omega_a$ for $a = 1, \dots,n-1$. The rotating-frame Hamiltonian is then 
\ba
\!\!\!\!\!\!\!\!\!\!\!\left(\!\!
\begin{array}{cccccc}
\Delta_1 & \Omega_1 & 0  & \cdots &0& 0\\
\Omega_1^* & \Delta_2 & \Omega_2 & \cdots &0& 0\\
0 & \Omega_2^* & \Delta_3 & \cdots & 0 & 0 \\
\vdots & \vdots & \vdots & \ddots & \vdots & \vdots \\
0 & 0 & 0 & \cdots & \Delta_{n-1} & \Omega_{n-1} \\
0 & 0 & 0 & \cdots & \Omega_{n-1}^* & \Delta_n
\end{array}
\!\!\!\right)\!\!
\left(\!\!\!
\begin{array}{c}
y_1\\
y_2\\
y_3\\
\vdots\\
y_{n-1}\\
y_n
\end{array}
\!\!\!\!\right) \!=\! E\! \left(\!\!\!
\begin{array}{c}
y_1\\
y_2\\
y_3\\
\vdots\\
y_{n-1}\\
y_n
\end{array}
\!\!\!\!\right). 
\ea
Here $\Delta_m = 0$ (recall that $\ket{m}$ is the reference sate), while $\Delta_{a+1}-\Delta_{a}$ is the detuning of the $a$'th microwave field from the $\ket{a}\rightarrow \ket{a+1}$ transition.
Writing
\ba
\Omega_a = \tilde \Omega_a \exp\left\{i \, \textrm{arg}\left(y_a/y_{a+1}\right)\right\}, 
\ea
where $\tilde \Omega_a$ are real, we obtain $n$ real-number equations 
\ba
\tilde \Omega_1 \left| \frac{y_2}{y_1}\right|&=& E - \Delta_1,\nonumber \\
\tilde \Omega_1 \left| \frac{y_1}{y_2}\right|   + \tilde \Omega_2  \left| \frac{y_3}{y_2}\right|  &=& E - \Delta_2,\nonumber \\
&\vdots&\nonumber \\
\tilde \Omega_{a-1}  \left| \frac{y_{a-1}}{y_a}\right| + \tilde \Omega_a  \left| \frac{y_{a+1}}{y_a}\right| &=& E - \Delta_a,\\
&\vdots&\nonumber \\
\tilde \Omega_{n-2}  \left| \frac{y_{n-2}}{y_{n-1}}\right|  + \tilde \Omega_{n-1}  \left| \frac{y_n}{y_{n-1}}\right|&=& E - \Delta_{n-1},\nonumber \\
\tilde \Omega_{n-1}   \left| \frac{y_{n-1}}{y_n}\right| &=& E - \Delta_{n}.\nonumber
\ea
One can satisfy the equations using $n-1$ Rabi frequencies $\tilde \Omega_a$ and any one nonzero detuning $\Delta_c$. Allowing other detunings to be nonzero, one obtains freedom in how to choose the magnitudes of $\tilde \Omega_a$. This can be useful, for example, in cases when the magnitudes of $\tilde \Omega_a$ need to be larger than a certain value (e.g.\ larger than the strength of hyperfine interactions, as in Refs.\ \cite{gorshkov11c,manmana12}  -- see our Outlook). 

The approach we have just described works nicely in the cases where the only off-diagonal bare-states processes that contribute to the final spin Hamiltonian have the form $\ket{a}\ket{b} \rightarrow \ket{b}\ket{a}$. This holds for all the configurations studied in Refs.\ \cite{gorshkov11b,gorshkov11c}. It also holds \footnote{The process $\ket{1,0} \ket{1,0} \rightarrow \ket{0,0} \ket{2,0}$ is not of the form $\ket{a}\ket{b} \rightarrow \ket{b}\ket{a}$ but the presented microwave construction still applies since we use the three bare states involved ($\ket{1,0}$, $\ket{1,1}$, and $\ket{1,2}$) as the reference states for the zero of the dressed-state energies.} 
for the level configuration used to implement the bilinear-biquadratic model in 
Ref.\ \cite{manmana12}. 

On the other hand, consider a model where $\{\ket{\u},\ket{\d}\} = \{\ket{0,0}, y_{-1} \ket{1,\!-\!1}+ y_1 \ket{1,1}\}$ (as in 
Ref.\ \cite{manmana12}) and where we want the process $\ket{1,\!-\!1} \ket{0,0} \rightarrow \ket{0,0} \ket{1,1}$ to contribute. We can couple states $\ket{1,\!-\!1}$ and $\ket{1,1}$ with two microwave fields via state $\ket{2,0}$. It is crucial that these two microwave fields have the same frequency; otherwise, in the rotating frame, the process $\ket{1,\!-\!1} \ket{0,0} \rightarrow \ket{0,0} \ket{1,1}$ would pick up an oscillatory time dependence and will average out to zero. This means that, in the cases where off-diagonal bare-state processes other than $\ket{a}\ket{b} \rightarrow \ket{b}\ket{a}$ contribute, there are restrictions on what detunings can be chosen as nonzero. 

In the case of the configuration for the Kitaev honeycomb model shown in Fig.\ \ref{fig:mw}(b), it is easy to check that the microwave couplings shown in Fig.\ \ref{fig:mw}(c) and Fig.\ \ref{fig:mw}(d) for $\ket{\u}$ and $\ket{\d}$, respectively, work. In particular, in Fig.\ \ref{fig:mw}(c) all microwaves are assumed to be resonant except for the microwaves coupled to state $\ket{2,0}$, which are all detuned by the same amount $\Delta$. This detuning provides the extra tuning parameter one has to add to the Rabi frequencies in order to fully control the coefficients $y_j$ defining state $\ket{\u}$. The fact that $\ket{2,0}$ has $M = 0$ ensures that it is nondegenerate. Hence, any contributing bare-state process containing $\ket{2,0}$ in the initial  two-molecule state will also contain it in the final two-molecule state, meaning that a detuning of this state keeps the process resonant. Similar arguments apply to Fig.\ \ref{fig:mw}(d), where all microwaves are assumed to be resonant except for the microwave coupling to state $\ket{4,0}$. To ensure that $\ket{\u}$ and $\ket{\d}$ are degenerate, we use $\ket{2,2}$ and $\ket{2,-2}$ as the two reference states and set the two dressed-state energies to be equal. It is also worth pointing out that an additional $N = 5$ state is introduced in both Fig.\ \ref{fig:mw}(c) and Fig.\ \ref{fig:mw}(d) to act as an intermediate state for coupling two states with $\Delta M = 2$. We also note that one has to be careful using $\pi$ transitions to couple states with $M \neq 0$ since the same microwave will act on both $M$ and $-M$. 

A solution alternative to the introduction of a detuning is an introduction of additional resonant microwave couplings, resulting in configurations such as three states coupled with three microwave fields. At the same time, it is worth pointing out that, in some cases, one may wish to introduce additional detunings on purpose with the goal of eliminating certain  interaction processes (written in terms of bare non-microwave-dressed states as $\ket{a,b} \rightarrow \ket{a',b'}$),  which can be regarded as an additional control knob.

\section{Outlook}

The most straightforward implementation of our proposal would involve polar molecules with no nuclear spin. This is possible, for example, in the case of a SrYb molecule, which seems to be within reach given the availability of quantum degenerate Sr and Yb gases \cite{daley11}. In the case where hyperfine structure is present, such as in alkali dimers, we assume that the Rabi frequencies $\Omega_a$ are larger than the strength of hyperfine interactions $H_\textrm{hf}$. In situations where this poses a significant restriction, the opposite limit, in which $\Omega_j \ll H_\textrm{hf}$, can also be considered.

While we have focused on pinned molecules, allowing the molecules to hop in the lattice gives rise to $t$--$J$-type \cite{gorshkov11b,gorshkov11c} or Hubbard-type \cite{barnett06}
 models with highly tunable long-range spin-spin interactions featuring a direction-dependent spin anisotropy. As a particularly simple example, 
 one can  obtain a single-component system with a density-density interaction whose angular dependence is given by a linear combination of all five $C^2_q$ [see the $\mathbf{V}$ term in Eq.\ (\ref{Hgen})]. This may allow one to study, for example, exotic quantum Hall physics \cite{qiu11}.

The approaches we presented can be used to  tune dipole-dipole interactions to zero \cite{buchler07b,gorshkov08c}. In that case, one has to consider  corrections due to dipole-dipole interactions to second order (which give van-der-Waals-type $1/R^6$ dependence) and interactions between higher multipole moments \cite{stone97}. These interactions can also be viewed as extra control knobs. 

We have focused in the present manuscript on polar molecules. However, the ideas that we presented should be extendable in a straightforward manner to other systems interacting via dipole-dipole interactions such as  magnetic atoms \cite{aikawa12,lu12}, Rydberg atoms \cite{saffman10,schauss12}, and magnetic solid-state defects \cite{dolde12,weber10}.



\section{Acknowledgments}

We thank S.\ Manmana, E.\ M.\ Stoudenmire, J.\ Preskill, J.\ Ye, D.\ Jin, M.\ Lukin, N.\ Yao, J.\ Taylor, S.\ Stellmer, W.\ Campbell, M.\ Foss-Feig, M.\ Hermele, and V.\ Gurarie for discussions. This work was supported by NSF, IQIM, NRC, AFOSR, ARO, 
the ARO-DARPA-OLE program, and  the Lee A. DuBridge and Gordon and Betty Moore foundations.  
KRAH thanks KITP for hospitality. 
This manuscript is the contribution of NIST and is not subject to U.S. copyright.

\end{document}